\documentclass[twocolumn,showpacs,preprintnumbers,amsmath,amssymb]{revtex4}

\usepackage{graphicx}
\usepackage{dcolumn}
\usepackage{bm}
\usepackage{color}
\usepackage{ulem}
\usepackage{natbib}
\usepackage{revsymb}
\usepackage{textcase}

\begin{document}

\title{Differential Emission Measure and Electron Distribution Function Reconstructed from RHESSI and SDO Observations}%

\author{G.G. Motorina$^{1,2}$}
 \affiliation{$^1$ Pulkovo Astronomical Observatory, Russian Academy of Sciences, Pulkovskoe sh. 65, St. Petersburg, 196140 Russian Federation; Email: g.motorina@yandex.ru}
\author{E.P. Kontar$^2$}%
\affiliation{$^2$ School of Physics and Astronomy, University of Glasgow, G12 8QQ, Glasgow, Scotland, United Kingdom
}

\received{February 10, 2015}

\begin{abstract}
\textbf{Abstract ---} To solve a number of problems in solar physics related to mechanisms of energy release in solar corona parameters of hot coronal plasma are required, such as energy distribution, emission measure, differential emission measure, and their evolution with time. Of special interest is the distribution of solar plasma by energies, which can evolve from a nearly Maxwellian distribution to a distribution with a more complex structure during a solar flare. The exact form of this distribution for low-energy particles, which receive the bulk of flare energy, is still poorly known; therefore, detailed investigations are required. We present a developed method of simultaneous fitting of data from two spacecrafts Solar Dynamics Observatory/Atmospheric Imaging Assembly (SDO/AIA) and Reuven Ramaty High Energy Solar Spectroscopic Imager (RHESSI), using a differential emission measure and a thin target model for the August 14, 2010 flare event.
\end{abstract}

\pacs{96.60.Q, 96.60.qe, 96.60.Tf}
\maketitle

\section{\label{sec:level1} Introduction}

Solar flares are magnetic explosive processes, spontaneously occurring in the solar atmosphere. The flares lead to an effective plasma heating and particle acceleration. Flare plasma temperature is normally diagnosed by studying of extreme ultraviolet radiation, while information about nonthermal plasma component (distribution of high-energy accelerated electrons) can be obtained from X-ray data. Solar flare observations with RHESSI [e.g. \onlinecite{2002SoPh..210....3L}] provide diagnostics on acceleration mechanisms, electron propagation \cite{2011SSRv..159..107H, 2011SSRv..159..301K} in the range from $3$ keV to $\sim 17$ MeV. New observations with SDO/AIA \cite{2012SoPh..275...17L} allowed to get images with high spatial (1.5$''$) and temporal ($\sim 12$ s) resolution, which made it possible to find a spatially-resolved differential emission measure  in a relatively wide temperature range (0.6-16 MK) \cite{2012A&A...539A.146H, 2013ApJ...771....2P, 2014ApJ...786...73S}.

Despite the extensive studies, the details of plasma heating, particle acceleration, and related processes are still poorly understood. Therefore, the use of newly available simultaneous observations allows to study hot plasma and energetic particles in flares in a wider energy range: for example, the EUV Variability Experiment (EVE) instrument of SDO and RHESSI \cite{2014ApJ...788L..31C} or SDO/AIA and RHESSI \cite{2013ApJ...779..107B,2014ApJ...789..116I}. The temperature range where SDO/AIA is more sensitive is approximately 0.6-16 MK, while RHESSI is more sensitive to temperatures above $\sim$10 MK. Thus, simultaneous observations in EUV and X-ray range provide a unique opportunity to study energy distribution of heated/accelerated electrons from $0.1$~keV up to several tens of keV in a solar flare.

In this paper, we develop and apply analytical functions suitable for both differential emission measure analysis and mean electron flux spectra in flares.

\subsection*{2.1. Connection of Differential Emission Measure with Distribution of Accelerated Electrons}

To obtain and analyze the spectrum of accelerated electrons, one should consider the differential emission measure (DEM), $\xi(T)$ [cm$^{-3}$~K$^{-1}$], i.e., the distribution of emitted plasma differential in temperature, which can be found from the expression [e.g. \onlinecite{2013ApJ...779..107B}]:
\begin{equation}
\langle nVF\rangle=\frac{2^{3/2}E}{\sqrt{\pi m_e}}\int_{0}^{\infty}\frac{\xi(T)}{(k_B T)^{3/2}}exp\Big(-\frac{E}{k_B T}\Big)dT,
\end{equation}
where $E$ is the electron kinetic energy, $m_e$ is the electron mass, $k_B$ is the Boltzmann constant, $\langle nVF\rangle$ is the mean electron flux spectrum [electrons keV$^{-1}$~s$^{-1}$~cm$^{-2}$].
Making the change of variables $t = 1/T$ in (1), one obtains
\begin{equation}
\langle nVF\rangle=\frac{2^{3/2}E}{\sqrt{\pi m_e}k_B^{3/2}}\int_{0}^{\infty}\frac{\xi(T(t))}{t^{1/2}}exp(-Et/k_B)dt.
\end{equation}

\begin{figure}
\includegraphics[width=8cm]{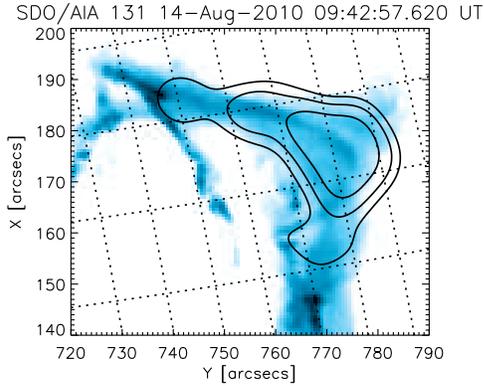}
\caption{AIA 131 \AA \; image with 20\%, 30\%, and 50\% RHESSI contours at 09:42-09:43 UT.}
\end{figure}

Based on Eq. (2), the differential emission measure was chosen in the way that the expression $\xi(T(t))/t^{1/2}$ has an analytical Laplace transform for further calculation of $\langle nVF\rangle$ and its behavior at low and high energies was similar to DEM obtained for SDO/AIA and RHESSI separately:
\begin{equation}
\xi(T)=\frac{EM}{T_0\Gamma(\alpha+1)}\Big(\frac{T}{T_0} \Big)^{\alpha}exp\Big(-\frac{T}{T_0} \Big),
\end{equation}
where $\Gamma(x)$ is the gamma function, $\alpha>0.5$: $\Gamma(x)=\int_{0}^{\infty}y^{x-1}exp(-y)dy$, which follows from DEM normalization: $\int_{0}^{\infty}\xi(T)dT=EM$. Thus substituting Eq. (3) into (2), $\langle nVF\rangle$ for DEM takes the form
\begin{eqnarray}
\langle nVF\rangle=\frac{2^{3/2}E}{\sqrt{\pi m_e}k_B^{3/2}}\frac{2EM(T_0E/k_B)^{0.5\alpha-0.25}}
{T_0^{\alpha+1}\Gamma(\alpha+1)}\nonumber \\
 \times K_{\alpha-0.5}\Big(2\sqrt{\frac{E/k_B}{T_0}} \Big),
\end{eqnarray}
where $K_n(z)$ is the modified Bessel function of the second kind \cite{Cody...1993}.

To find DEM from observations, it is necessary to vary three parameters: $EM$, $T_0$, and $\alpha$. Although parameter $T_0$ has no specific physical meaning,
this parameter can always be recalculated through the maximum
temperature $T_{max}=\alpha T_0$, which corresponds to a maximum
of the function $\xi(T)$ or through the average temperature $\langle T\rangle=EM^{-1}\int_{0}^{\infty}T \xi(T)dT$, which is $\langle T\rangle=T_0(\alpha+1)$ in our case. Thus we can conclude that the fitting
method allows to obtain not only the analytical form of DEM and $\langle nVF\rangle$ but also (automatically) the key plasma parameters that make it possible to diagnose plasma over wide range of temperatures.

\subsection*{2.2. The August 14, 2010 Solar Flare}

We consider the solar flare of August 14, 2010 \cite{2013ApJ...779..107B}, which started at 09:25:40~UT and refers to a GOES C4.1 class flare \cite{2005SoPh..227..231W}. The flare was well observed with both RHESSI and SDO/AIA.

The RHESSI soft X-ray data were taken at 09:42-09:43~UT before the flare peak at 09:46 UT. Using fitting in OSPEX of the RHESSI data with a multi-thermal model $multi\_therm\_2pow$:
\begin{equation}
\xi(T)=\frac{EM}{T_0B(\alpha+1,\beta-\alpha-1)}\Big(\frac{T}{T_0} \Big)^{\alpha}\Big(1+\frac{T}{T_0} \Big)^{-\beta},
\end{equation}
where $B(x, y)$ is the beta-function:
$B(x, y)=\int_{0}^{\infty}\frac{t^{x-1}}{(1+t)^{x+y}}dt$ and using a thin target model $thin2$ [for example, see \onlinecite{1988psf..book.....T}] with $\chi^2=0.84$ the following parameters were obtained: $EM=5\times10^{47}$ cm$^{-–3}$, $T_0$ = 0.75 keV, $T_{max}=T_0\alpha/(\beta-\alpha)$ = 0.25 keV, $\alpha$=3, $\beta$=12, the spectral index $\delta$ = 3.2, and the low-energy cut-off $E_c=7.75$ keV. For the same time interval GOES temperature and emission measure were $T_{GOES}$ = 0.8 keV
and $EM_{GOES}=5\times10^{47}$ cm$^{-3}$.

The data in EUV range obtained from SDO/AIA in six EUV filters (131 \AA (Fe VIII, Fe XX, Fe XXIII), 171 \AA (Fe IX), 193 \AA (FeXII, FeXXIV), 211 \AA (Fe XIV), 335 \AA (Fe XVI), and 94 \AA (Fe X, Fe XVIII)) were additionally calibrated in terms of  translation, rotation, and scaling using the program $aia\_prep.pro$ and normalized to the exposure time. The errors on SDO/AIA data ($DN$) included the systematic error and were calculated by the formula $DN_{err}=(DN+(0.2DN)^2)^{1/2}$. It should be noted that the SDO/AIA images were taken at almost the same time and the time interval between them did not exceed 12 s. It was assumed that the same emitting plasma is observed in all wavelengths from the volume corresponding to 50$\%$ RHESSI contour. Figure 1 shows the AIA 131 \AA\; image (09:42:57.62 UT) with RHESSI 20$\%$, 30$\%$, and 50$\%$ contours for the energy range
of $8-10$~keV, CLEAN algorithm \cite{2002SoPh..210...61H} for the time interval 09:42-09:43 UT. Thus the SDO/AIA data from the looptop out of the region corresponding to 50$\%$ RHESSI contour were used to find DEM.
\begin{figure*}
\includegraphics[width=9.5cm]{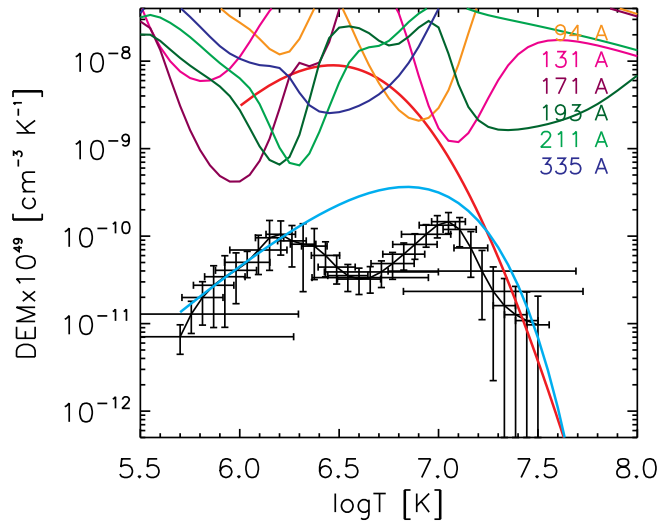}
\caption{\label{fig:epsart}DEMs: regularized DEM from SDO/AIA data (black solid line), DEM from RHESSI data (red line), DEM from combined SDO/AIA and RHESSI data (blue line), the remaining lines are SDO/AIA loci-curves.}
\end{figure*}

\begin{figure*}
\includegraphics[width=14cm]{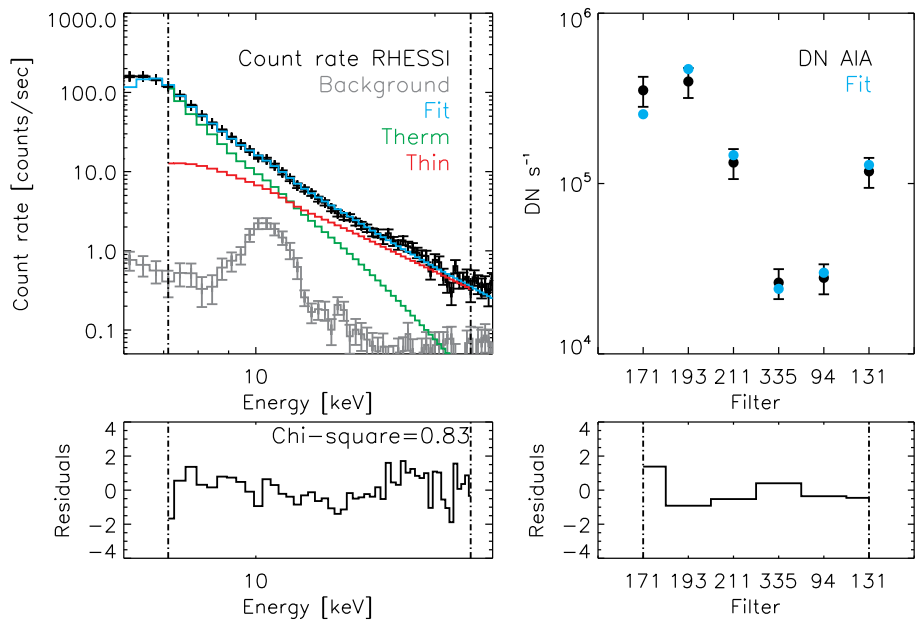}
\caption{\label{fig:epsart} Results of simultaneous fitting of RHESSI (left) and SDO/AIA (right) data. (Top left) RHESSI data (Count rate RHESSI), fitting results (Fit), thermal (Therm) and nonthermal (Thin) models; (top right): SDO/AIA data (DN AIA, black circles), fitting results (Fit, blue circles); (bottom) the ratio of the difference between observed and fitted data to the corresponding errors for RHESSI and SDO/AIA measurements.}
\end{figure*}

\subsection*{2.3. Inference of Differential Emission Measure from SDO/AIA and RHESSI Data}

 Since each SDO/AIA passband is sensitive to different temperatures, each SDO/AIA passband provides an additional data point for use in the fitting method. We determine DEM parameters from a small number of  SDO/AIA passbands and a number of RHESSI energy bins.

For the given area (Fig. 1), we find DEM by three different methods:

(1) the regularization method \cite{Tikh...1979, 2004SoPh..225..293K, 2005SoPh..226..317K,2012A&A...539A.146H,2012JTePh..57.1618M} for the SDO/AIA data \cite{2013SoPh..283....5A};

(2) OSPEX fitting with a multi-thermal function with DEM in the form of Eq. (5) for the RHESSI data;

(3) fitting with a thermal model where DEM is represented as Eq. (2) and a nonthermal model (thin-target) simultaneously for RHESSI and SDO/AIA data.

Figure 2 shows the differential emission measure for three methods and SDO/AIA loci-curves.

It can be seen from Fig. 2 that DEM has a complex structure, therefore it is rather difficult to find a simple functional form. Also we should note that the attempts to fit the data with only a single DEM function without adding a nonthermal model turned out to be unsuccessful for the given flare ($\chi^2>1$).
This indicates that the choice of DEM functions themselves is unsuccessful or that there is a nonthermal component.

\begin{figure}
\includegraphics[width=7cm]{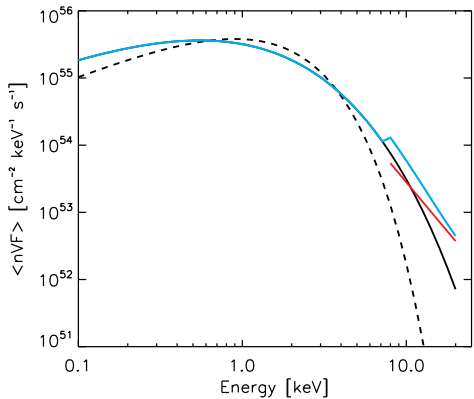}
\caption{Mean electron flux spectrum found from fitting of SDO/AIA and RHESSI data with DEM function (black solid line) and a thin target (red line), and their sum (blue line), as well as the Maxwellian distribution corresponding to $EM = 4.64\times 10^{46}$ cm$^{-–3}$ and $T=\langle T\rangle=0.9$ keV (black dashed line).}
\end{figure}

Using a simultaneous fitting of RHESSI and SDO/AIA data sets ($\chi^2=0.83$), we have the following parameters: $EM=4.64\times10^{46}$ cm$^{-–3}$, $T_0$ = 0.3 keV, $T_{max}$ = 0.58 keV, $\langle T\rangle$ = 0.9 keV, $\alpha$ = 1.96, the spectral index is $\delta$ = 2.9, and the low-energy cut-off $E_c=7.78$ keV. Figure 3 shows the fitting results. In the figure, the AIA filters were ordered to show the temperature increase at which the filters are most sensitive. Figure 4 shows the mean electron flux spectrum $\langle nVF\rangle$ (obtained from Eq. (4)) for the resulting parameters and the Maxwellian distribution corresponding to the values of the resulting emission measure and the average temperature.

 The residuals show that the fitting closely matches the data from combination of RHESSI and SDO/AIA observations. However, it should be noted that the fit is not good enough for a filter with a wavelength of 171 \AA\; which is responsible for low temperatures.

\section*{3. CONCLUSIONS}

In this study we have developed a method for finding the differential
emission measure as an functional form that fits both RHESSI and SDO/AIA data. Using this method we reconstructed the differential emission measure and the mean electron flux spectrum for August 14, 2010 flare event. It has been shown that the DEM obtained from RHESSI and SDO/AIA observations can be fitted using a simple analytical form, which can be presented via the mean electron flux spectrum of electrons in the flare. The difference between SDO/AIA data and the fitting results for the filter 171 \AA\; is about one sigma, which can be explained by the fact that the line of sight intersects the plasma with a temperature of $\sim 10^6$ K located above or below the coronal loop for which the DEM is calculated.

Using a combined analysis of SDO/AIA and RHESSI data we found for the first time the mean electron flux spectrum for a wide energy range (0.1-20 keV). It has been shown that the deviation of $\langle nVF\rangle$ from the Maxwellian distribution is present not only at high but also low energies, i.e., the distribution of particles has a more complex structure.

\section*{ACKNOWLEDGMENTS}
This study was supported by programs P-9 and P-41 of the Presidium of the Russian Academy of Sciences, RFBR grants 13-02-00277A and 14-02-00924A, and the Marie Curie International Research Staff Exchange Scheme ”Radiosun” (PEOPLE-2011-IRSES-295272).

\bibliographystyle{apsrev}
\bibliography{apssamp1}

\end{document}